\documentclass[preprint,aps,nofootinbib]{revtex4}
\usepackage{graphicx}
\usepackage{epsfig}
\usepackage{amsmath}
\usepackage{amsfonts}
\usepackage{amssymb}
\usepackage{color}
\usepackage{dcolumn}
\usepackage{multirow}
\usepackage{booktabs}
\usepackage{subfigure}
\usepackage{CJKutf8}
\usepackage{mathrsfs}
\usepackage{graphics}
\usepackage{float}
\usepackage{indentfirst}
\usepackage{hyperref}
\providecommand{\U}[1]{\protect\rule{.1in}{.1in}}
\newcolumntype{L}{>{$}l<{$}}
\newcolumntype{R}{>{$}r<{$}}

\newcommand{\f}{\begin{equation}}
\newcommand{\ff}{\end{equation}}
\newcommand{\fa}{\begin{eqnarray}}
\newcommand{\ffa}{\end{eqnarray}}

\begin{document}
\title{Correspondence between grey-body factors and quasinormal modes for regular black holes with sub-Planckian curvature}
\author{Chen Tang$^{1}$}
\email{ctangphys@stu.cwnu.edu.cn} 
\author{Yi Ling $^{2,3,1}$}
\email{lingy@ihep.ac.cn(Corresponding author)}
\author{Qing-Quan Jiang$^{1}$}
\email{qqjiangphys@yeah.net(Corresponding author)}
 \affiliation{$^1$ School of Physics and Astronomy, China West Normal University, Nanchong 
637002, China \\$^2$ Institute of High Energy
Physics, Chinese Academy of Sciences, Beijing 100049, China\\$^3$
School of Physics, University of Chinese Academy of Sciences,
Beijing 100049, China}
\begin{abstract}  
We investigate the quasi-normal modes (QNMs) under the gravitational field perturbations and the grey-body factors for a class of regular black holes with sub-Planckian curvature and Minkowski core. Specifically, we compute the QNMs with the pseudospectral method and the WKB method. It is found that as the deviation parameter of the regular black hole changes, the trajectory of the QNMs displays non-monotonic behavior. Then  we compute the grey-body factors with the WKB method and compare them with the results obtained by the correspondence relation recently revealed in \cite{Konoplya:2024lir}. We find the discrepancy exhibits minimor errors, indicating that this relation is effective for computing  the grey-body factors of such regular black holes. 
\end{abstract}
\maketitle
\section{Introduction}

The Hawking radiation of a black hole actually traverses the gravitational field surrounding the black hole, which modifies the spectrum of the radiation and renders it distinct from the ideal black-body radiation\cite{hawking1975particle,page1976particle,visser1998hawking}.
The grey-body factor accounts for the discrepancy between the actual spectrum of black hole radiation and the ideal case of perfect black-body radiation, and it modifies the purely theoretical prediction of the Hawking radiation spectrum. It is a frequency-dependent function that characterizes the probability that radiation at a given frequency can traverse the potential barrier surrounding the black hole and escape to infinity. Due to the strong gravitational field of a black hole, part of the radiation is reflected back into the black hole or absorbed by it, resulting in a weaker observed radiation intensity. Therefore, the grey-body factor is pivotal in refining the description of Hawking radiation, bringing theoretical calculations closer to the actual physical conditions
\cite{page1976particle,sanchez1978absorption,Maldacena:1996ix,Cvetic:1997uw,visser1998hawking,Kanti:2002ge,Boonserm:2008zg,Pantig:2022gih,lan2021quasinormal,Ovgun:2024zmt,Okabayashi:2024qbz}.

On the other hand, the QNMs as the response of a black hole under the perturbations can be used to reveal the intrinsic characteristics of black holes and compact stars, such as the mass and spin\cite{Nollert:1999ji,kokkotas1999quasi,berti2009quasinormal,konoplya2011quasinormal,flanagan1998measuring,yang2012quasinormal}. At the same time, it is an important tool for analyzing gravitational waves during the ringdown process\cite{berti2009quasinormal,Berti:2016lat,abbott2016observation,cardoso2016gravitational,Baibhav:2017jhs,Giesler:2019uxc,JimenezForteza:2020cve,Maggio:2020jml,Dey:2020pth,Carullo:2021oxn,rosato2024ringdown,Yi:2024elj}. Recently, the QNMs have also been widely studied as an effective means of probing the internal structure and quantum effects of regular black holes\cite{berti2006gravitational,bronnikov2012instabilities,flachi2013quasinormal,jusufi2021quasinormal,lan2021quasinormal,hendi2021physical,konoplya2023quasinormal,meng2023gravito,lopez2023quasi,gingrich2024quasinormal,konoplya2024infinite,franchini2024testing,zhang2024quasinormal,Pedrotti:2024znu,Tang:2024txx}. Regular black holes represent a special class of black hole models characterized by the absence of a singularity at their center\cite{bardeen1968non,dymnikova1992vacuum,hayward2006formation,frolov2014information}. Unlike standard black holes in general relativity, regular black holes circumvent the singularity problem, thereby avoiding the physical conundrum of infinite density and curvature at the core of traditional black holes\cite{Penrose:1964wq,ch5-}. The concept of regular black holes was initially proposed based on phenomenological considerations, suggesting that quantum gravitational effects would eliminate the singularity under extreme conditions, resulting in a more “regular” internal structure of the black hole\cite{Xiang:2013sza,Li:2016yfd,Bianchi:2018mml,Simpson:2018tsi}.
Subsequently, it is shown that regular black holes can be constructed in various ways. One approach involves solving the Einstein field equations with exotic matter that violates classical energy conditions, thereby obtaining black hole solutions without singularities\cite{ayon1998regular,balart2014regular,koch2014black,fan2016construction,ayon2000bardeen}. Another approach is to construct regular black hole solutions by considering modified Einstein equations that incorporate quantum gravitational corrections, which typically arise from quantum gravity theories or semi-classical gravitational theories\cite{Ashtekar:2018cay,Ashtekar:2018lag,achour2018polymer,Feng:2024sdo}.
 The study of regular black holes not only provides new insights into resolving the singularity problem of traditional black holes but also offers a significant theoretical framework for exploring the manifestations of quantum gravitational effects in strong gravitational fields.

As a matter of fact, both the QNMs and the grey-body factors are two important quantities that describe the properties of black holes with different boundary conditions. Interestingly, in  \cite{Konoplya:2024lir}, the correspondence between the QNMs and the grey-body factor is established with the help of WKB expansion. Specifically, a formula is derived that allows one to compute the grey-body factor from the fundamental mode and the first overtone of the QNMs for large values of the angular number $l$. In particular, in the eikonal limit the grey-body factor is solely determined by the fundamental mode. 
Currently, this correspondence in the high-frequency regime has been extended to spherically symmetric black holes\cite{Konoplya:2024lir}, axisymmetric black holes\cite{Konoplya:2024vuj}, quantum-corrected black holes\cite{skvortsova2024quantum}, and other types of black holes\cite{Malik:2024cgb,Malik:2024wvs,Bolokhov:2024otn,Dubinsky:2024vbn}. In this paper we intend to construct this correspondence for  a sort of regular black hole proposed in \cite{ling2023regular}. These regular black holes are characterized by sub-Planckian curvature and a Minkowski core, unlike the ordinary Bardeen and Hayward black holes that possess a de-Sitter core.

Our work is organized as follows. In Section \ref{Regular black holes metrics and  WAVE-LIKE EQUATION}, we briefly introduce regular black holes with sub-Planckian curvature and a Minkowskian core. In Section \ref{QUASINORMAL MODES OF THE REGULAR BLACK HOLE}, we compute the QNMs of regular black holes under gravitational perturbations. In Section \ref{GREY-BODY FACTORS OBTAINED VIA THE CORRESPONDENCE WITH QUASINORMAL MODES}, we focus on the greybody factors and numerically check  the correspondence between the QNMs and grey-body factors. Section \ref{CONCLUSION AND DISCUSSION} is the  summary and conclusions.
\section{Regular black holes metrics and  WAVE-LIKE EQUATION}
\label{Regular black holes metrics and  WAVE-LIKE EQUATION}

\begin{figure}[ht]
		\centering{
			\includegraphics[width=7.8cm]{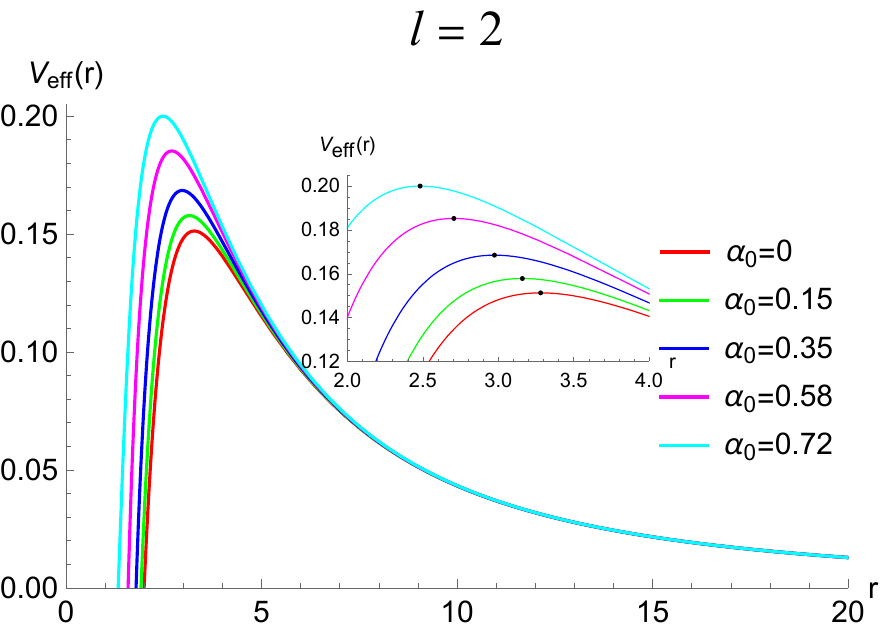}\hspace{6mm}
			\includegraphics[width=7.8cm]{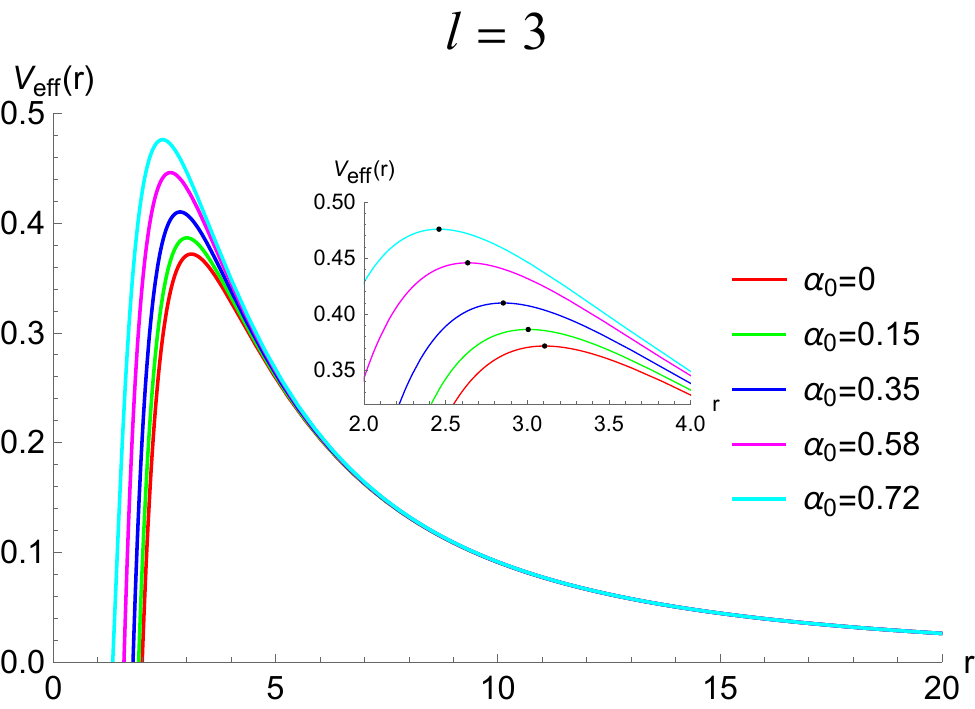}
			\caption{The effective potential $V_{eff}(r)$ of the gravitational field with different deviation parameters $\alpha_0$. The black dot represents the maximum value of the $V_{{eff}}(r)$.}
			\label{veff}
		}
\end{figure}

In this section, we briefly introduce the regular black hole which was originally proposed in \cite{ling2023regular}. This kind of regular black hole is characterized by the sub-Planckian curvature as well as an asymptotically Minkowski core, which is realized  by introducing an exponentially suppressing
gravity potential \cite{Xiang:2013sza}. Specifically, the spherically symmetric metric of this sort of regular black holes takes the form as follows
\begin{equation} 
    d s^{2}=-f(r) d t^{2}+\frac{1}{f(r)}d r^{2}+r^{2}\left(d \theta^{2}+\sin ^{2} \theta d \phi^{2}\right),
    \label{metric}
\end{equation}
where the function $f(r)$ is given by
\begin{equation} 
    f(r)=1+2\psi(r)=1-\frac{2M}re^{\frac{-\alpha_0M^x}{r^c}},
    \label{psi}
\end{equation}
where $M$ is understood as the mass of black hole and the parameters $x$, $c$, $\alpha_0$ are all the dimensionless parameters \footnote{We have set $8\pi G=l_p^2=1$ throughout this paper for convenience. After recovering the dimension of the potential, it manifestly becomes $\psi(r)=-\frac{MG}re^{\frac{-\alpha_0(MG)^xl_p^{c-x}}{r^c}}$.}. Here the value of $\alpha_0$ reflects the degree of deviation from the Newtonian potential and characterizes the corrections due to the effects of quantum gravity\cite{Xiang:2013sza,Li:2016yfd,ling2023regular}. The theoretical properties and the observation signature of this sort of regular black holes have been recently investigated in \cite{Ling:2022vrv,zeng2023astrophysical,Zeng:2023fqy,zhang2024quasinormal,Tang:2024txx,Guo:2025zca}. Without loss of generality, in this paper we will firstly focus on the case of $x=2/3$ and $c=2$, which exhibits the same asymptotical behavior as Bardeen black hole at infinity, and then present a discussion on other cases with different values of $x$ and $c$. In this case, the Hawking temperature is given by $f'(r_h)/(4\pi)$, and to keep its positivity one finds that the value of $\alpha_0$ should satisfy
\begin{equation} 
	\alpha_0\leq\frac{2M^\frac{4}{3}}{e}.
 \label{alpha}
\end{equation}
Next we consider the QNMs of the gravitational field perturbations over this regular black hole. Previously, the perturbations of the scalar field with spin zero as well as the electromagnetic field with spin one have been investigated in detail in \cite{Tang:2024txx}. By means of the separation of variables, the axial perturbation equation can be simplified to a Schrödinger-like equation :
\begin{equation}\label{Sch_like_eq}
   \frac{\partial^2\Psi}{\partial r_*^2}+(\omega^2-V_{eff})\Psi=0\:,
\end{equation}
where the tortoise coordinate $r_*$ is defined in relation to the radial coordinate $r$ by the equation:
\begin{equation}
 \frac{dr_*}{dr} = \frac{1}{f(r)}.   
\end{equation}
The effective potential under gravitational perturbations has the form
\begin{equation}
 V_{eff}(r) = f(r) \left( \frac{l(l + 1)}{r^2} - \frac{3 f'(r)}{r} \right),   
\end{equation}
where $l$ is the angular number. In Fig. \ref{veff}, we plot the effective potential $V_{eff}(r)$ of the
gravitational field with different deviation parameters $\alpha_0$. It shows that the effective potential is always positive, indicating that the spacetime of this regular BH is stable under the gravitational perturbation. 
From Fig. \ref{veff}, we note two characteristics of the effective potential. First, when $l$ is fixed, the peak of the effective potential increases as the deviation parameter $\alpha_0$ increases. Second, when $\alpha_0$ is fixed, the maximum of the effective potential increases as $l$ increases.
\section{QUASINORMAL MODES OF THE REGULAR BLACK HOLE}
\label{QUASINORMAL MODES OF THE REGULAR BLACK HOLE}
In this section, we investigate the QNMs of regular black holes with $x=2/3$ and $c=2$ under gravitational perturbations. It is well known that the boundary conditions for QNM are defined as follows:
\begin{subequations}
\begin{align}
   \Psi &\sim e^{-i\omega(t+r_*)}\! & r_* &\to +\infty, \\
     \Psi &\sim e^{-i\omega(t-r_*)}\! & r_* &\to -\infty.
\end{align}
\end{subequations}
The frequencies of QNMs can be determined by solving the wave equation in conjunction with the boundary conditions.
In this section, we employ the WKB method and the pseudospectral method to compute the QNMs of the regular black hole under gravitational perturbations. The details of the pseudospectral method can be found in the appendix of \cite{Tang:2024txx}. Here we present the numerical results as below. 

Tables \ref{n=0,1 l=2} and \ref{n=0,1 l=3} list the QNMs of $n=0$ and $n=1$ for different values of the deviation parameter with using the WKB method and the pseudospectral method, respectively. The discrepancy between these two methods is minimal, which validates the accuracy of both approaches. For $n=0$ and $n=1$, the QNMs exhibit a larger real part and a smaller imaginary part, indicating stronger oscillations and weaker damping in both cases. For different values of $\alpha_0$, with the same overtone number $n$ and angular number $l$, the QNMs frequency changes in a monotonic manner as $\alpha_0$ increases. Specifically, the real part of the QNMs frequency increases, while the imaginary part decreases.

\begin{table}
\centering
\begin{tabular}{c|cc|cc} 
\hline
           & \multicolumn{2}{c|}{$n=0$}                    & \multicolumn{2}{c}{$n=1$}                      \\ 
\hline
$\alpha_0$ & \multicolumn{1}{c|}{WKB} & PS                 & \multicolumn{1}{c|}{WKB} & PS                  \\ 
\hline
0          & 0.373619-0.088891I       & 0.373672-0.088962I & 0.346297-0.27348I        & 0.346711-0.273915I  \\ 
\hline
0.15       & 0.381319-0.085311I       & 0.382617-0.08763I  & 0.352538-0.264144I       & 0.3586-0.269413I    \\ 
\hline
0.3        & 0.39348-0.085922I       & 0.393707-0.085653I & 0.3724-0.264948I         & 0.37302-0.262788I   \\ 
\hline
0.5        & 0.410334-0.08219I       & 0.411063-0.08163I  & 0.390747-0.254688I       & 0.394524-0.249416I  \\ 
\hline
2/e        & 0.441664-0.069059I       & 0.437952-0.071494I & 0.437082-0.196175I       & 0.420167-0.217681I  \\
\hline
\end{tabular}
\caption{The fundamental mode and the first overtone mode  under gravitational perturbations for different values of $\alpha_0$, where $l=2$. }
\label{n=0,1 l=2}
\end{table}

\begin{table}
\centering
\begin{tabular}{c|cc|cc} 
\hline
           & \multicolumn{2}{c|}{$n=0$}                     & \multicolumn{2}{c}{$n=1$}                       \\ 
\hline
$\alpha_0$ & \multicolumn{1}{c|}{WKB} & PS                  & \multicolumn{1}{c|}{WKB} & PS                   \\ 
\hline
0          & 0.599443-0.092703I     & 0.599443-0.092703I & 0.582642-0.28129I      & 0.582644-0.281298I  \\ 
\hline
0.15       & 0.611912-0.091489I       & 0.611485-0.091532I  & 0.596558-0.277408I   & 0.596085-0.277538I   \\ 
\hline
0.3        & 0.62612-0.089774I        & 0.62623-0.089753I   & 0.612282-0.27194 I      & 0.612408-0.27186I    \\ 
\hline
0.5        & 0.648871-0.086103I      & 0.648869-0.086098I  & 0.636886-0.260279I      & 0.636885-0.260262I   \\ 
\hline
2/e        & 0.684759-0.076612I       & 0.683071-0.077217I  & 0.670626-0.231287I      & 0.669319-0.233063I   \\
\hline
\end{tabular}
\caption{The fundamental mode and the first overtone mode  under gravitational perturbations for different values of $\alpha_0$, where $l=3$.}
\label{n=0,1 l=3}
\end{table}

\begin{table}
\centering
\begin{tabular}{c|c|c|c|c} 
\hline
$\alpha_0$ & $n=2$                & $n=3$                 & $n=4$                & $n=5$                \\ 
\hline
0     & 0.301053-0.478277I & 0.251505-0.705148I  & 0.207515-0.946845I & 0.169303-1.19561I  \\ 
\hline
0.15  & 0.318382-0.468975I & 0.275574-0.689117I  & 0.239356-0.923063I & 0.211068-1.163552I  \\ 
\hline
0.3   & 0.338577-0.455611I & 0.302014-0.666708I  & 0.271387-0.890632I & 0.248115-1.121092I  \\ 
\hline
0.5   & 0.366331-0.429253I & 0.334302-0.623543I  & 0.303987-0.829115I & 0.276284-1.041561I  \\ 
\hline
2/e   & 0.384833-0.374182I & 0.334921-0.549445I & 0.279652-0.749675I & 0.23127-0.972254I  \\
\hline
\end{tabular}
\caption{higher overtone  modes under gravitational perturbations for different values of $\alpha_0$ , where $l=2$}
\label{n=2,3,4,5 l=2}
\end{table}

\begin{table}
\centering
\begin{tabular}{c|c|c|c|c} 
\hline
$\alpha_0$ & $n=2$                & $n=3$                 & $n=4$                & $n=5$                \\ 
\hline
0     & 0.551685-0.479093I &0.511962-0.690337I  & 0.470174-0.915649I & 0.431386-1.15215I  \\ 
\hline
0.15  & 0.567742-0.471985I & 0.531359-0.678713I  & 0.492955-0.898341I  &0.457169-1.128414I  \\ 
\hline
0.3   & 0.586941-0.461381I & 0.554056-0.661643I & 0.518891-0.873288I & 0.485507-1.094386I  \\ 
\hline
0.5   & 0.614486-0.440002I & 0.584606-0.627843I  &0.550755-0.824515I & 0.515777-1.02902I  \\ 
\hline
2/e   &0.642013-0.393333I &0.601909-0.561527I& 0.551089-0.741569I & 0.493761-0.936819I  \\
\hline
\end{tabular}
\caption{higher overtone  modes under gravitational perturbations for different values of $\alpha_0$ , where $l=3$}
\label{n=2,3,4,5 l=3}
\end{table}

Next we turn to the higher overtone  modes. It is important to emphasize that the WKB method exhibits high precision only when the overtone number $n$ is relatively small. Therefore, for data where $n \geq 2$
, we employ the pseudospectral method for the calculations. From Tables \ref{n=2,3,4,5 l=2} and \ref{n=2,3,4,5 l=3}, it can be seen that higher overtone QNMs have smaller real parts and larger imaginary parts which is in contrast to the case of $n=0$ and $n=1$, implying that these overtone modes are damping rapidly. Furthermore, for a given $l$ and $\alpha_0$, the real part of the frequency becomes smaller as the overtone number 
$n$ increases, while the imaginary part becomes larger. On the other hand, if we fix $\alpha_0$ and $n$, but increase $l$, then from Tables \ref{n=2,3,4,5 l=2} ($l=2$) and \ref{n=2,3,4,5 l=3}  ($l=3$), we find that the real part of the QNMs increases as well. This is understandable because as $l$ increases, the effective potential increases as well  (refer to Fig. \ref{veff}), implying that a larger frequency is required to overcome the potential barrier. Finally, when we examine the effects of the variation of $\alpha_0$, we find for low overtone numbers, the real and imaginary part of the QNMs change monotonically with increasing $\alpha_0$; however, for high overtone numbers, this monotonic behavior disappears.  
To gain an intuitive picture for the influence of $\alpha_0$ on QNMs, we demonstrate the change of the QNM frequency with the deviation parameter $\alpha_0$ in Fig. \ref{l=2,n=0，1，2，3，4，5}  for $l=2$  and Fig. \ref{l=3,n=0，1，2，3，4，5} for $l=3$, respectively. In each plot with a fixed $n$ and $l$, the variation of the frequency with $\alpha_0$ forms a trajectory on the frequency plane as $\alpha_0$ runs from $0$ to $2/e$, where the left ending point corresponds to $\alpha_0=0$ (representing the Schwarzschild black hole), while the right ending point represents $\alpha_0=2/e$. 

\begin{figure}[ht]
    \centering
    \includegraphics[width=5.2cm]{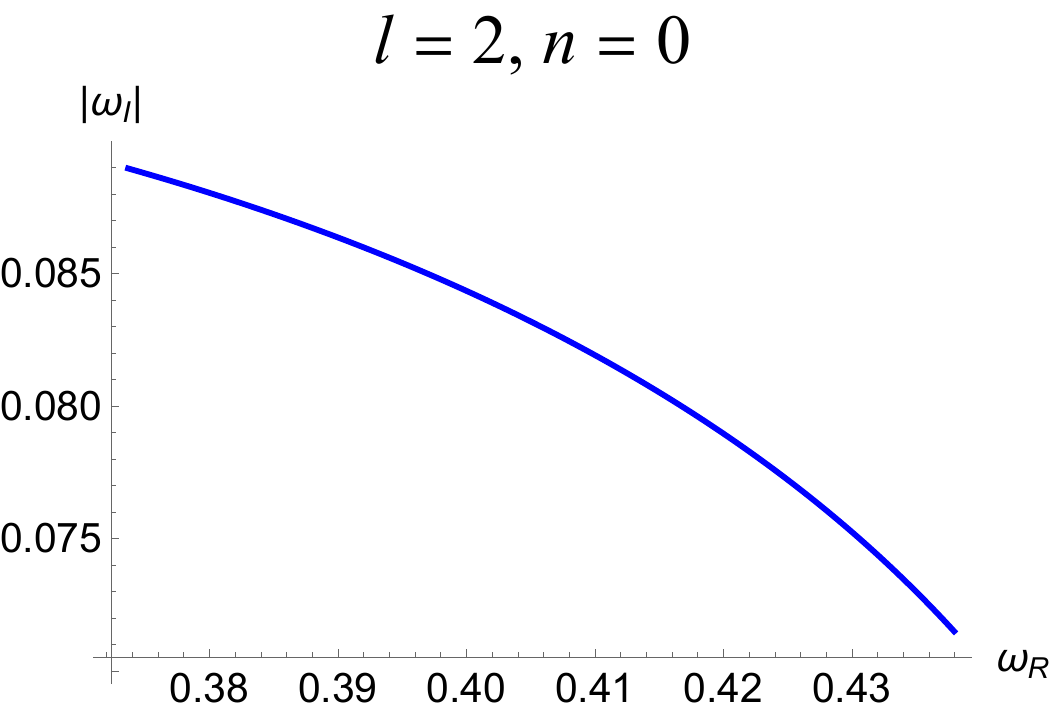}\hspace{1mm}
    \includegraphics[width=5.2cm]{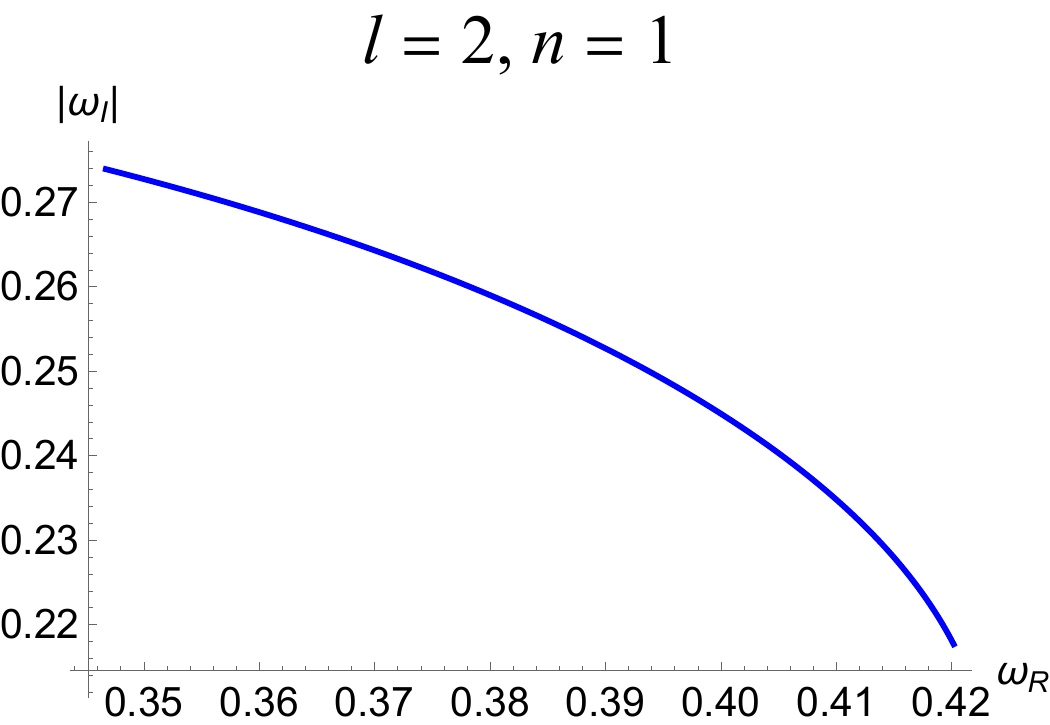}\hspace{1mm}
    \includegraphics[width=5.2cm]{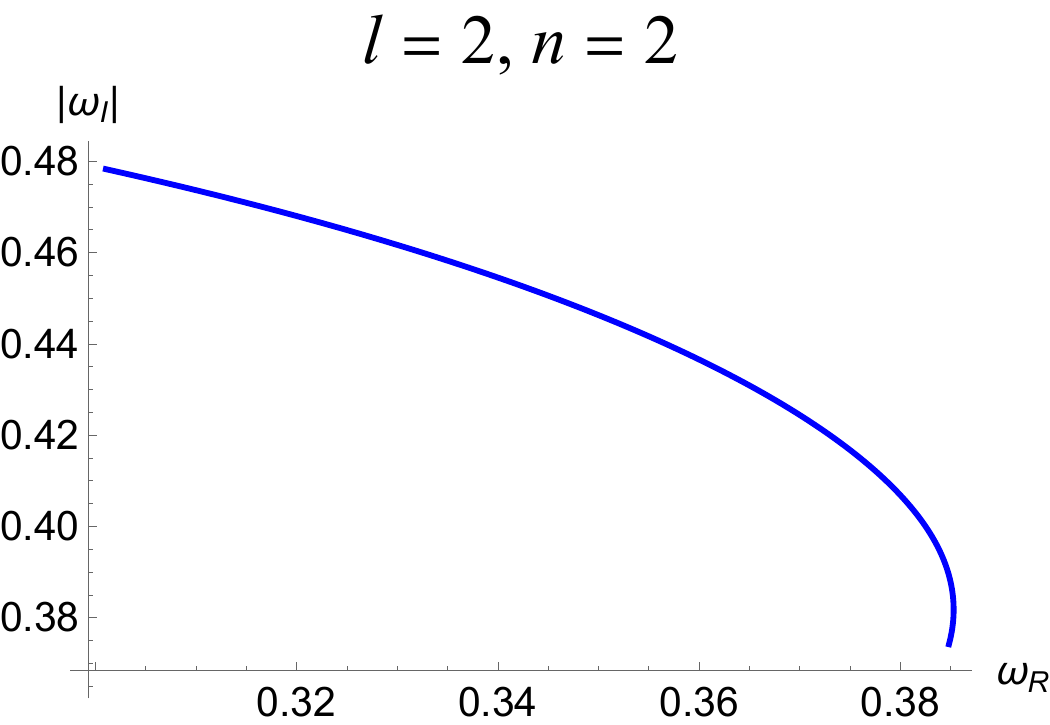}\hspace{1mm}
    \includegraphics[width=5.2cm]{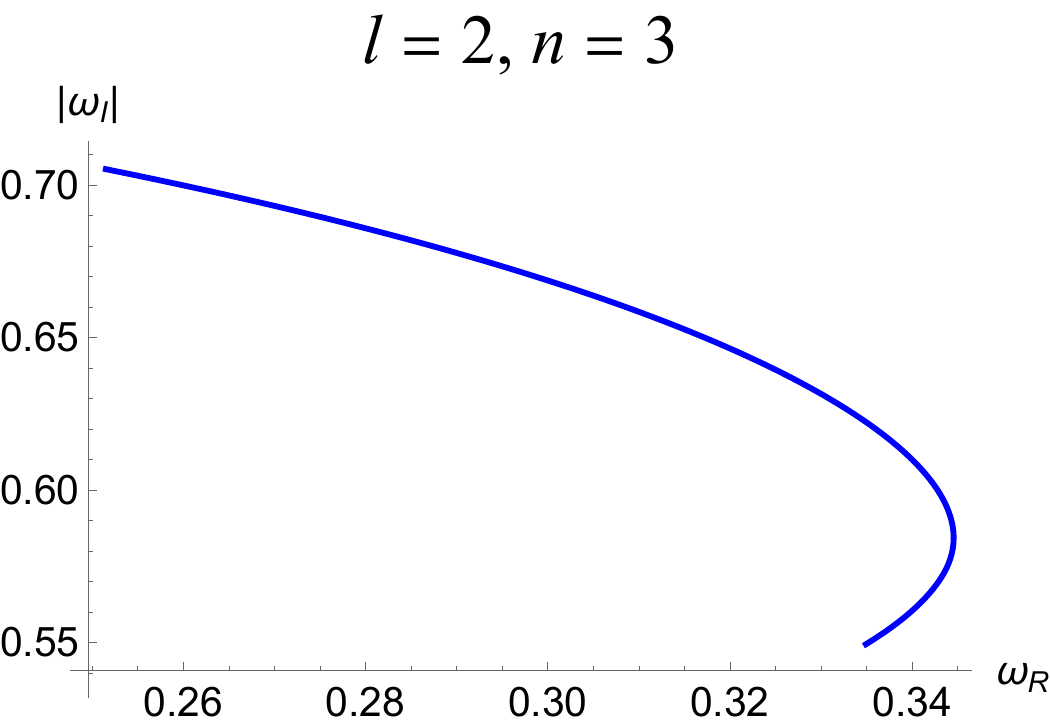}\hspace{1mm}
    \includegraphics[width=5.2cm]{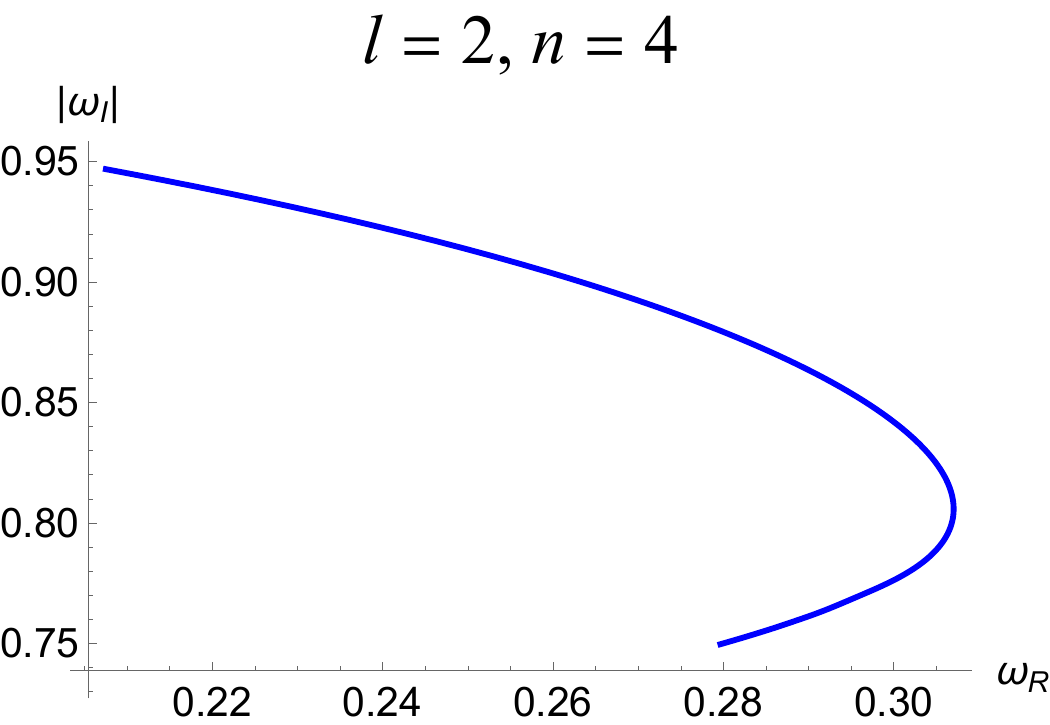}\hspace{1mm}
    \includegraphics[width=5.2cm]{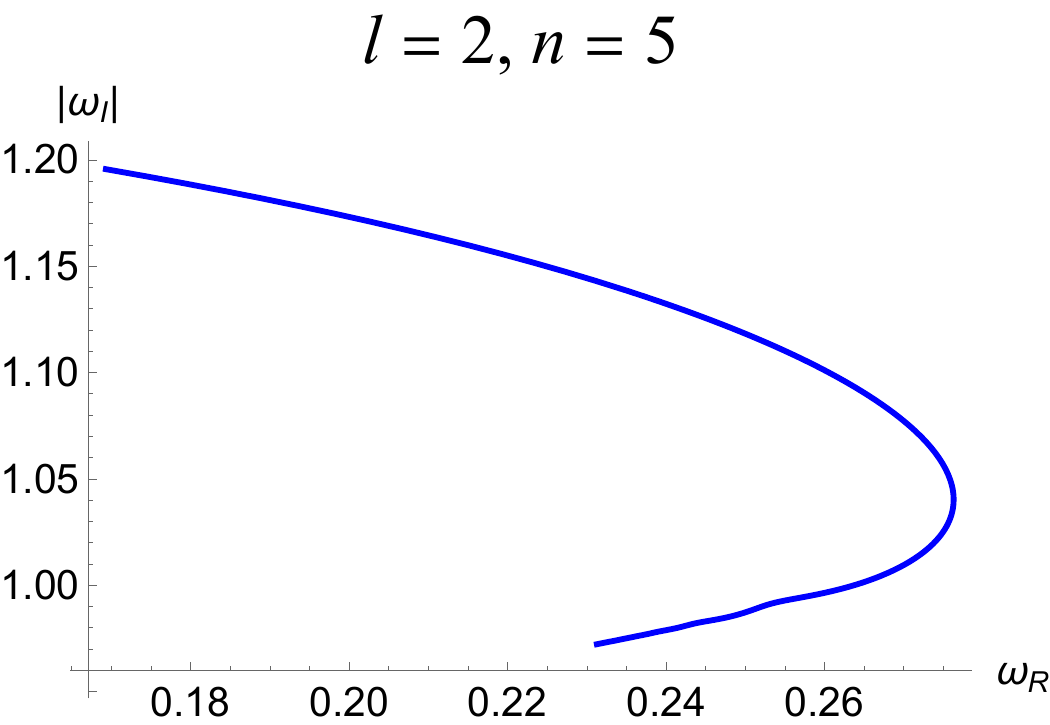}\hspace{1mm}
    \caption{The trajectory of the QNMs with the variation of $\alpha_0$ on the frequency plane with $l=2$ under gravitational field perturbations.}
		\label{l=2,n=0，1，2，3，4，5}
\end{figure}

\begin{figure}[ht]
    \centering
    \includegraphics[width=5.2cm]{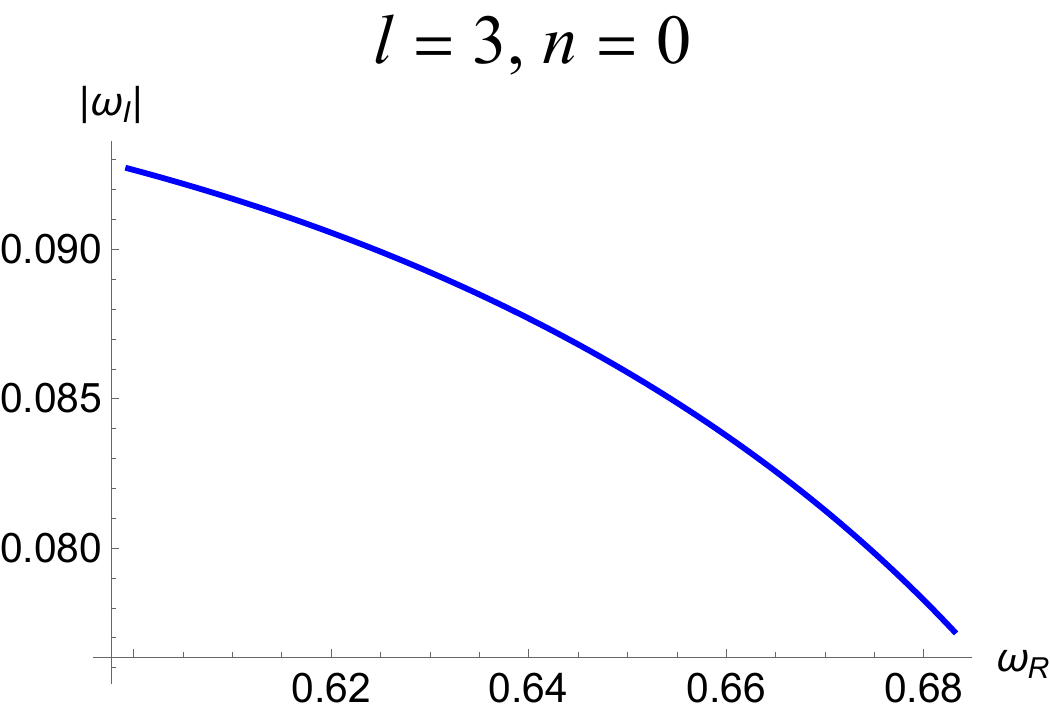}\hspace{1mm}
    \includegraphics[width=5.2cm]{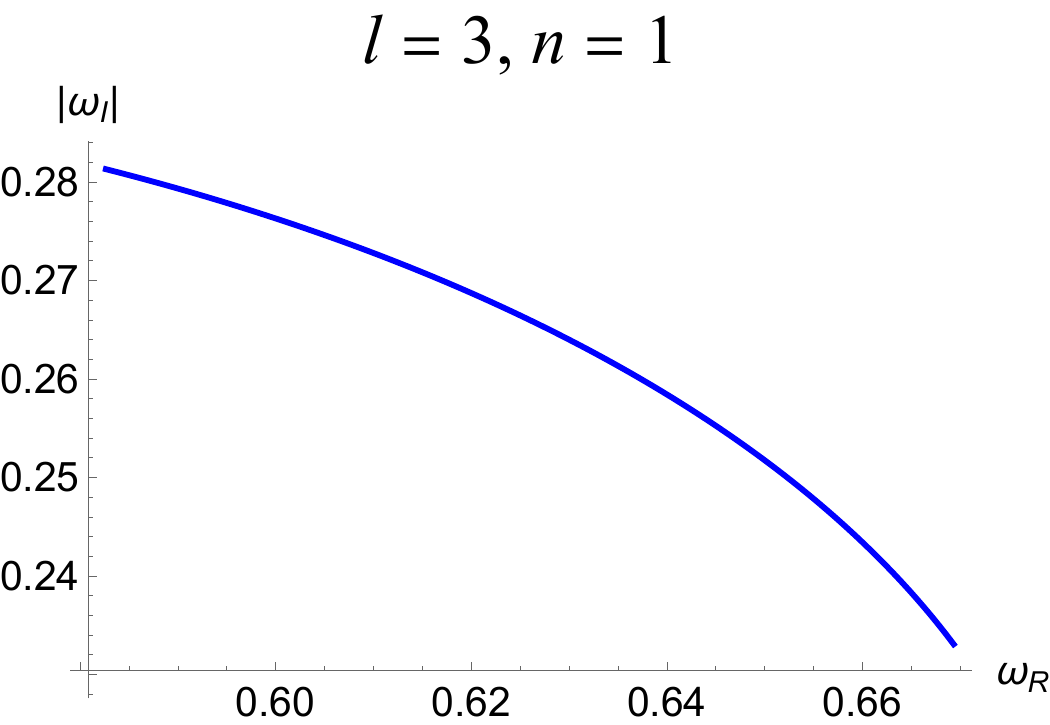}\hspace{1mm}
    \includegraphics[width=5.2cm]{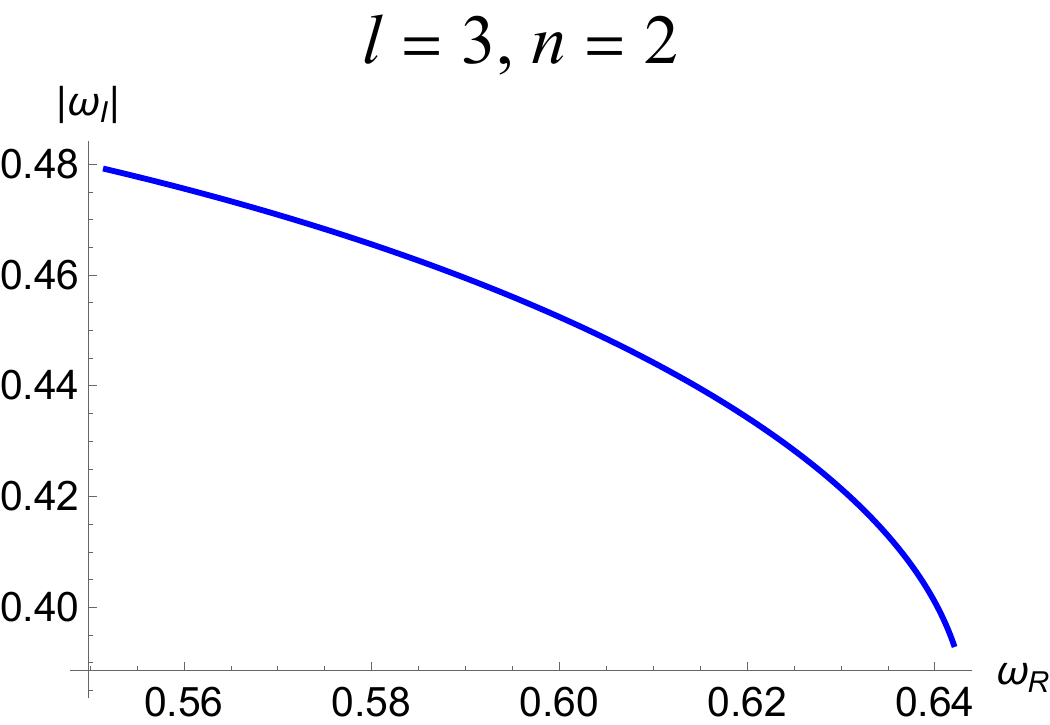}\hspace{1mm}
    \includegraphics[width=5.2cm]{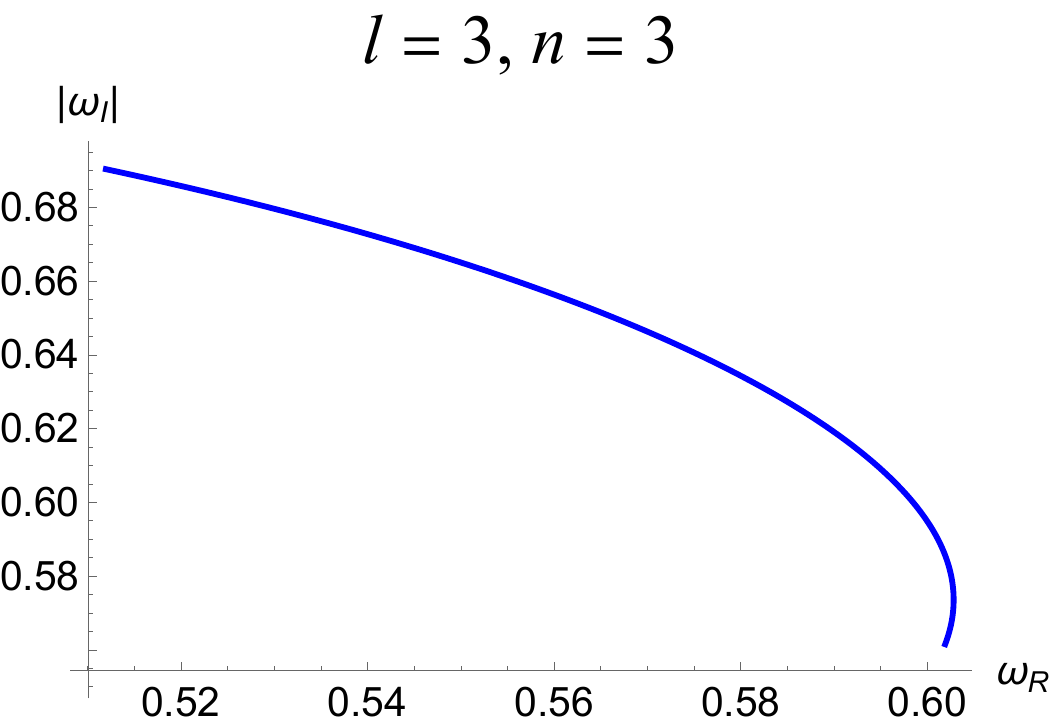}\hspace{1mm}
    \includegraphics[width=5.2cm]{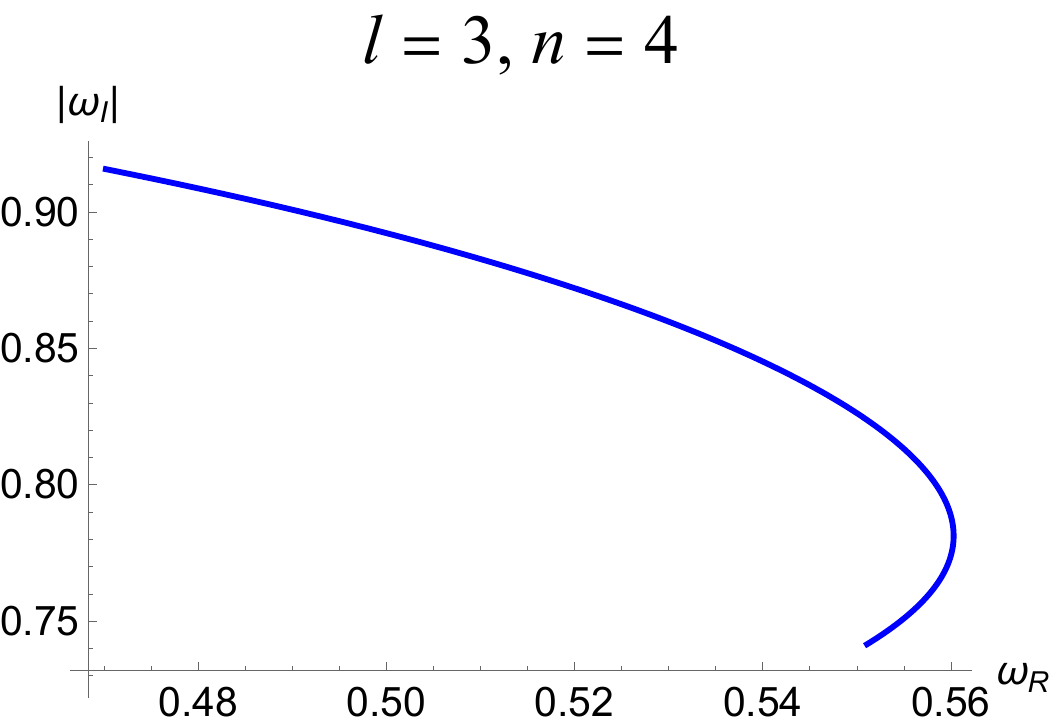}\hspace{1mm}
    \includegraphics[width=5.2cm]{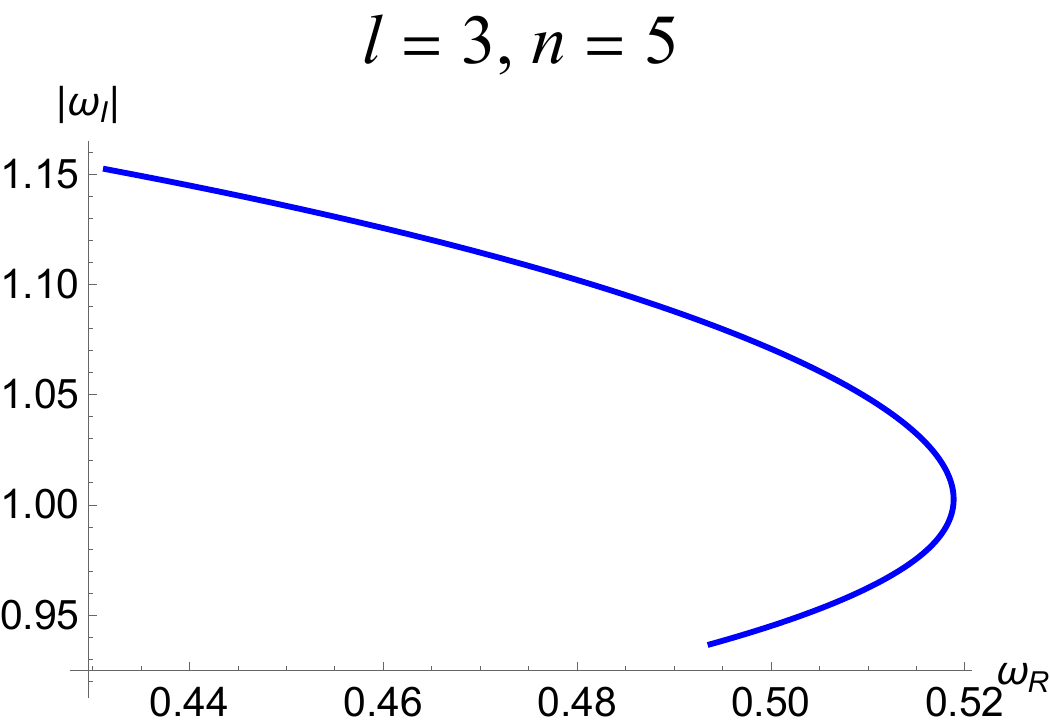}\hspace{1mm}
    \caption{The trajectory of the QNMs with the variation of $\alpha_0$ on the frequency plane with $l=3$} under gravitational field perturbation.
		\label{l=3,n=0，1，2，3，4，5}
\end{figure}

From Figs. \ref{l=2,n=0，1，2，3，4，5} and \ref{l=3,n=0，1，2，3，4，5}, it is observed that, overall, as $\alpha_0$ increases, the real part of the QNMs increases while the imaginary part decreases. This monotonic behavior persists for $l=2, n=0,1 $ and $l=3, n=0,1,2$. For higher overtone numbers, as  $\alpha_0$  increases, the real part of the QNMs first increases and then decreases, while the imaginary part decreases. This non-monotonic behavior illustrated in figures is quite similar to the results in  \cite{zhang2024quasinormal,Tang:2024txx} regarding the trajectories of QNMs for electromagnetic and scalar perturbations. The difference lies in the fact that in references \cite{zhang2024quasinormal,Tang:2024txx}, when $n$ is large, the QNM trajectories not only exhibit non-monotonicity but also develop a spiral structure for $l=0$ and $l=1$. However, in the case of gravitational perturbations studied here, we have $l\ge 2$ such a spiral structure does not appear. It is anticipated that increasing the overtone numbers $n$ might lead to this interesting spiral behavior.

\begin{figure}[ht]
		\centering{
			\includegraphics[width=7.8cm]{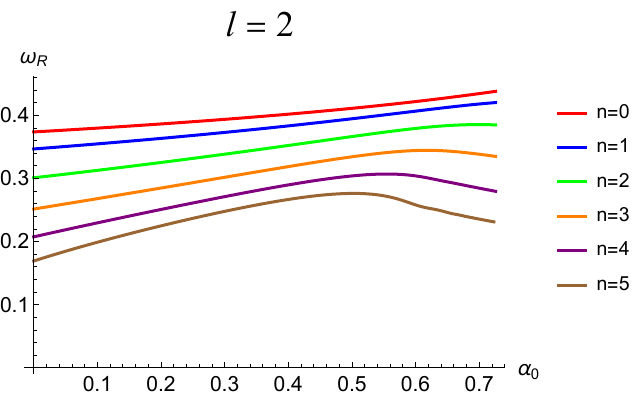}\hspace{6mm}
			\includegraphics[width=7.8cm]{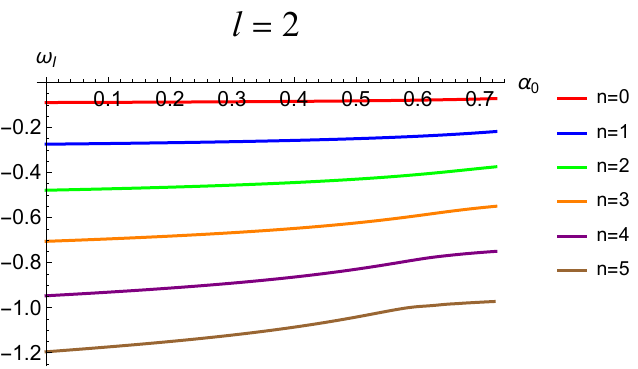}
			\caption{The real and imaginary parts of QNMs as the function of $\alpha_0$ with $l=2$ under gravitational field perturbation.}
			\label{Real part Imaginary part l=2}
		}
\end{figure}

\begin{figure}[ht]
		\centering{
			\includegraphics[width=7.8cm]{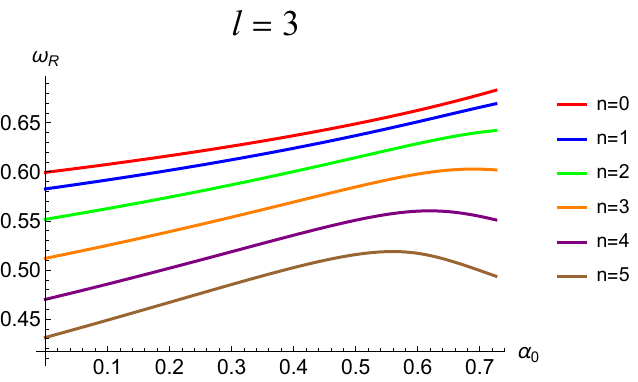}\hspace{6mm}
			\includegraphics[width=7.8cm]{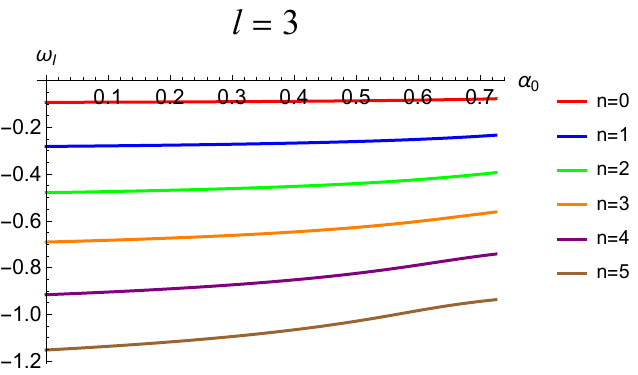}
			\caption{The real and imaginary parts of QNMs as the function of $\alpha_0$ with $l=3$ gravitational field perturbation.}
			\label{Real part Imaginary part l=3}
		}
\end{figure}

Figs. \ref{Real part Imaginary part l=2} and \ref{Real part Imaginary part l=3} illustrate the real and imaginary parts of the QNMs as 
a function of the deviation parameter $\alpha_0$ separately, providing another angle of view on their changes. In these figures,
it is evident that when $n$ increases to larger values, the real part first significantly increases and then decreases, while the imaginary part also undergoes some minor changes. In comparison with the results obtained in \cite{zhang2024quasinormal,Tang:2024txx}, where the separation of the real and imaginary parts of the QNMs is performed for the scalar fields, the oscillatory behavior does not appear here. In future studies, if a spiral structure emerges with increasing $\alpha_0$ and $n$, then it is expected that the oscillatory behavior would appear in the separated real and imaginary parts.

\section{GREY-BODY FACTORS OBTAINED VIA THE CORRESPONDENCE WITH QUASINORMAL MODES}
\label{GREY-BODY FACTORS OBTAINED VIA THE CORRESPONDENCE WITH QUASINORMAL MODES}

In this section we turn to the grey-body factor of this sort of regular black holes. We will compute it with the use of the sixth-order WKB method and then compare the results obtained via the correspondence relation between the QNMs and the grey-body factor. We will focus on  the regular black hole with $(x=2/3, c=2)$, and then present a brief discussion on the other regular black holes with different values of $x$ and $c$.  

The grey-body factor scattering has the following boundary conditions:
\begin{equation}
\begin{aligned}&\Phi=Te^{-i\Omega r_{*}} ,\quad r_{*}\to-\infty ,\\&\Phi=e^{-i\Omega r_{*}}+Re^{i\Omega r_{*}} ,\quad r_{*}\to\infty,\end{aligned}
\end{equation}
where $T$ is the transmission coefficient and $R$ is the reflection coefficient. Here, $\Omega$ is a real number, in contrast to QNMs, which have complex frequencies. According to normalization, the transmission coefficient $T$ and the reflection coefficient $R$ are related by the following equation:
\begin{equation}
|T|^2+|R|^2=1.
\end{equation}
The reflection coefficient $R$ is approximately calculated using the WKB method as follows:
\begin{equation}
R=(1+e^{-2i\pi\mathcal K})^{-1/2},
\end{equation}
where $\mathcal{K}$ is determined by the following formula:
\begin{equation}
\mathcal{K}=i\frac{\omega^2-V_0}{\sqrt{-2V_2}}-\sum_{k=2}^{k=6}\Lambda_k(\mathcal{K}),
\end{equation}
where $\Lambda_k(\mathcal{K})$ is the higher WKB corrections \cite{schutz1985black,konoplya2003quasinormal,konoplya2019higher}. Thus, the grey-body factor for different values of $l$ is calculated as:
\begin{equation}
|A_l|^2=1-|R|^2=|T|^2.
\end{equation}

The correspondence between the QNMs and the grey-body factor has been established in \cite{Konoplya:2024lir},  and it is described by an analytical expansion in terms of the angular number $l$ with the following form:
\begin{equation}
\begin{aligned}
i\mathcal{K} & =\frac{\Omega^2-Re(\omega_0)^2}{4Re(\omega_0)Im(\omega_0)}\left(1+\frac{(Re(\omega_0)-Re(\omega_1))^2}{32Im(\omega_0)^2}-\frac{3Im(\omega_0)-Im(\omega_1)}{24Im(\omega_0)}\right) \\
 & -\frac{Re(\omega_0)-Re(\omega_1)}{16Im(\omega_0)}-\frac{(\omega^2-Re(\omega_0)^2)^2}{16Re(\omega_0)^3Im(\omega_0)}\left(1+\frac{Re(\omega_0)(Re(\omega_0)-Re(\omega_1))}{4Im(\omega_0)^2}\right) \\
 & +\frac{(\omega^2-Re(\omega_0)^2)^3}{32Re(\omega_0)^5Im(\omega_0)}\left(1+\frac{Re(\omega_0)(Re(\omega_0)-Re(\omega_1))}{4Im(\omega_0)^2}\right. \\
 & +Re(\omega_0)^2\left(\frac{(Re(\omega_0)-Re(\omega_1))^2}{16Im(\omega_0)^4}-\left.\frac{3Im(\omega_0)-Im(\omega_1)}{12Im(\omega_0)}\right)\right)+\mathcal{O}\left(\frac{1}{\ell^3}\right),
\end{aligned}
\end{equation}
 where $\Omega$ represents the frequency of the grey-body factor, while $\omega_0$ and $\omega_1$ denote the fundamental mode and the first overtone mode of the QNMs, respectively.

\begin{figure}[ht]
	\centering{
		\includegraphics[width=7.8cm]{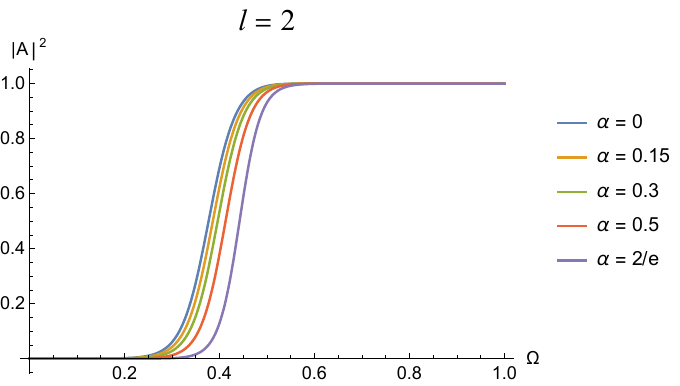}\hspace{6mm}
		\raisebox{4mm}{\includegraphics[width=7.8cm]{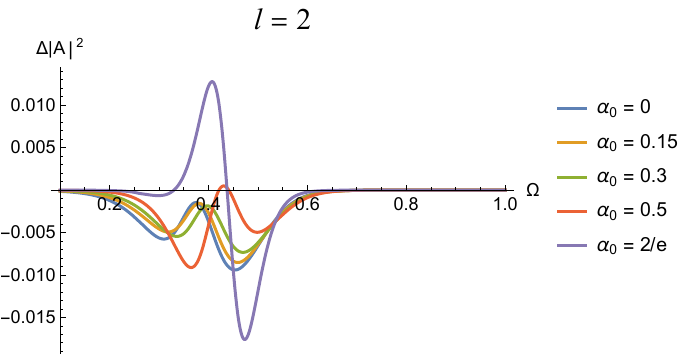}}\hspace{6mm}
		\includegraphics[width=7.8cm]{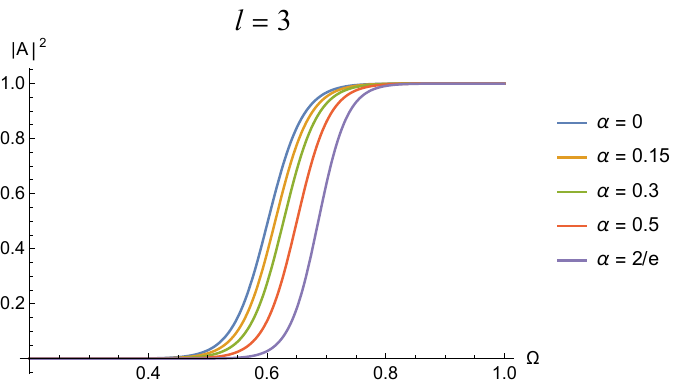}\hspace{6mm}
		\raisebox{4mm}{\includegraphics[width=7.8cm]{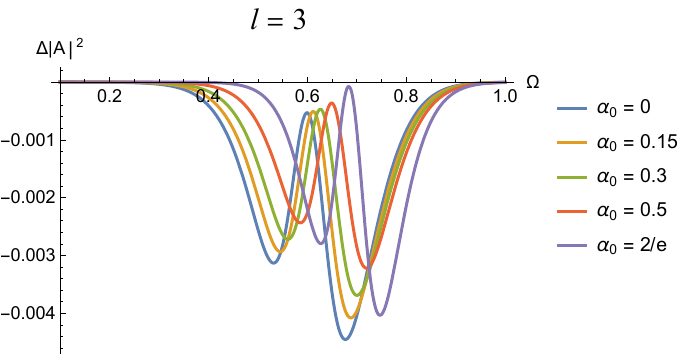}}\hspace{6mm}
		\includegraphics[width=7.8cm]{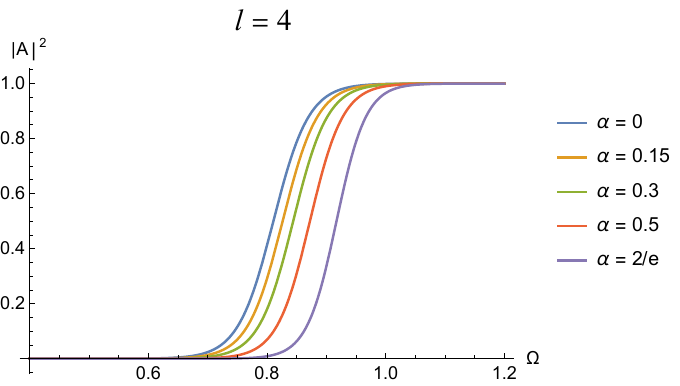}\hspace{6mm}
		\raisebox{4mm}{\includegraphics[width=7.8cm]{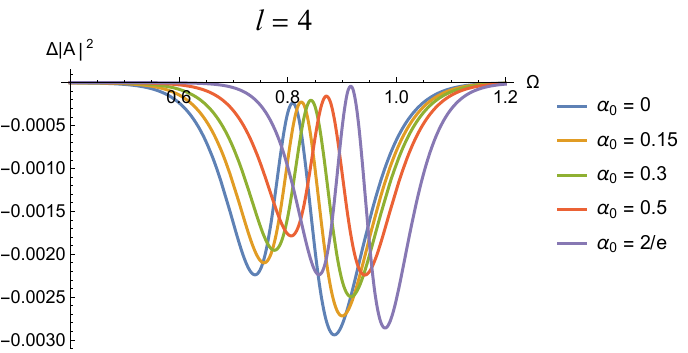}}\hspace{6mm}
		\caption{Left image: the grey-body factors derived from the correspondence of regular black holes with $l=2,3,4$ for different deviation parameters $\alpha_0$. Right image: the discrepancy between the results by the correspondence relation and those by sixth-order WKB method.}
		\label{grey body l=2,3,4}
	}
\end{figure}

Now we intend to check this correspondence for the regular black holes with $x=2/3$ and $c=2$. Fig. \ref{grey body l=2,3,4} illustrates the grey-body factor as the function of the frequency $\Omega$ and the difference between the grey-body factors obtained through the correspondence relation and those obtained using the sixth-order WKB method. From these figures, we have following observations and remarks: 
\begin{itemize}
    \item Regarding the variation of grey-body factor with $\Omega$: With the increase of the frequency $\Omega$, the grey-body factor gradually increases from 0 to 1. This indicates that as the energy of particles becomes larger,
    they are more likely to penetrate the potential barrier, resulting in a larger transmission coefficient and thus an increase in the grey-body factor.
    \item  Regarding the variation of grey-body factor with $\alpha_0$: As the deviation parameter $\alpha_0$ increases, the frequency required to achieve the same grey-body factor also increases. This is because an increase in $\alpha_0$ leads to a higher effective potential, requiring particles to have a higher frequency to penetrate the barrier.
    \item Regarding the difference in grey-body factors: From the right plots, it can be seen that the difference between the grey-body factors obtained through the correspondence relation and those obtained using the sixth-order WKB approximation is very small, and this difference exhibits a similar wave-like pattern. Notably, as $l$ increases from 2 to 4, the difference between the grey-body factors further decreases, with the discrepancy being only one-thousandth. This further confirms the accuracy of the grey-body factors obtained through the correspondence relation, and it is expected that this correspondence becomes exact in the limit of $l\rightarrow \infty$.
\end{itemize} 

To check the universality of this correspondence, we present a brief discussion on other regular black holes with different values of $x$ and $c$ in the end of this section. Without loss of generality, we consider $x = 1$ and $c = 3$. This type of regular black hole has the same asymptotical behavior as the Hayward black hole at large scales. With the same algebra, one can derive that 
the deviation parameter $\alpha_0$ for this regular black hole ranges from 0 to 8/3e.
The QNMs of this black hole under the scalar perturbations were investigated in  
\cite{zhang2024quasinormal}. Here we present the QNMs of regular black hole with $x=1$ and $c=3$ under the gravitational perturbations in Table \ref{QNMHayward}. In particular, we list the discrepancy of the grey-body factors obtained by the correspondence relation and the sixth-order WKB method separately, where $\Delta |A|^2$ denotes the maximal value of the discrepancy as it runs as the function of the frequency $\Omega$. The results show that the correspondence relation is also valid for this regular black hole, and the discrepancy decreases as the angular number $l$ increases.

\begin{table}
\centering
\begin{tabular}{c|c|c|c|c|c|c} 
\hline
                    & \multicolumn{3}{c|}{$l=2$}                 & \multicolumn{3}{c}{$l=3$}                    \\ 
\hline
$\alpha_0$          & $n=0$              & $n=1$              &  $\Delta |A|^2$ & $n=0$               & $n=1$              &  $\Delta |A|^2$ \\ 
\hline
0 & 0.373672-0.088962I & 0.346711-0.273915I & 0.0094 & 0.599443 -0.092703I & 0.582644-0.281298I & 0.0045  \\ 
\hline
0.25& 0.373672-0.088962I & 0.356405-0.265287I & 0.0087 & 0.607122 -0.090469I & 0.591806-0.274238I & 0.0041 \\ 
\hline
0.5 & 0.386371-0.083093I & 0.366047-0.254532I & 0.0088 & 0.615229-0.087722I  & 0.601026-0.265527I & 0.0035  \\ 
\hline
0.75 & 0.393932-0.078508I & 0.375307-0.239352I &0.019& 0.624505-0.083924I & 0.610438-0.25354I & 0.0036 \\ 
\hline
8\textbackslash{}3e & 0.401236-0.072533I & 0.379621-0.221364I & 0.05 & 0.63369-0.079133I & 0.617221-0.23921I &   0.0056 \\
\hline
\end{tabular}
\caption{The QNMs of regular black hole ($x=1,c=3$) under gravitational perturbations for different values of $\alpha_0$, with $n=0,1$ and $l=2,3$. The grey-body factors are obtained by the correspondence relation and the sixth-order WKB method separately, with $\Delta |A|^2$ denoting the maximal value of the discrepancy as it runs as the function of the frequency $\Omega$.}
\label{QNMHayward}
\end{table}

\section{CONCLUSION AND DISCUSSION}
\label{CONCLUSION AND DISCUSSION}
In this paper we have computed the QNMs of regular black holes with sub-Planckian curvature under gravitational perturbations by employing the pseudospectral method and the WKB method. As the overtone numbers $n$ increases, it is observed that the trajectory of the QNMs  with the variation of the deviation parameter $\alpha_0$ exhibits non-monotonic behavior on the frequency plane. Since for gravitational perturbations $l\ge 2$, we did not observe the spiral behavior of the QNM over the frequency plane as revealed  in \cite{zhang2024quasinormal,Tang:2024txx}, where the QNMs for the scalar fields and the vector fields were investigated and the spiral behavior is observed for $l=0$ and $l=1$ only. Thus we intend to conclude that the increase of the angular number $l$ would suppress the non-monotonic and spiral behavior of the QNMs on the frequency plane. On the other hand, it is more likely to observe the non-monotonic behavior and spiral behavior of the frequency for larger overtone number $n$, leading to the outburst of the overtone. We expect to justify this statement by computing the QNMs for larger $n$ in future.  
Furthermore, we have computed  the grey-body factors with the WKB method and then compared them with the results obtained with the correspondence relation between the QNMs and the grey-body factors. 
The results indicate that for small $l$, the grey-body factors derived from the correspondence relation have tiny difference compared to those from the WKB method, and these discrepancy decrease further as the angular number $l$ increases, demonstrating the reliability of this approach for regular black holes with sub-Planckian curvature and Minkowski core. 

\section*{Acknowledgments}
We are grateful to Guo-ping Li, Kai Li, Wen-bin Pan, Meng-he Wu, and Zhangping Yu for helpful discussions on the QNMs and the relevant numerical methods. This work is supported in part by the Central Guidance on Local Science and Technology Development Fund of Sichuan Province (2024ZYD0075) , and by the Natural Science Foundation of China (Grant Nos.~12035016,~12275275)
\bibliographystyle{style1}
\bibliography{grey}

\end{document}